\documentclass{appolb}
\usepackage{graphicx}
\usepackage{amsmath,amssymb}
\usepackage{hyperref}
\usepackage{cite}

\newcommand{\as}{\alpha_s}

\begin{document}
\title{Small-$x$ resummation in PDF fits and future prospects%
\thanks{Presented at ``Diffraction and Low-$x$ 2022'', Corigliano Calabro (Italy), September 24-30, 2022.}%
}
\author{Marco Bonvini
\address{INFN, Sezione di Roma 1,\\ Piazzale Aldo Moro~5, 00185 Roma, Italy}
}
\maketitle
\begin{abstract}
I review recent progress in the determination of PDFs with the inclusion of small-$x$ resummation,
and its impact in precision phenomenology, and discuss future prospects.
\end{abstract}
  
\section{PDF fits with small-$x$ resummation}

The resummation of high-energy (small-$x$) logarithms in QCD processes has been
an active field of research for more than 30 years. In the context of QCD collinear factorization,
which is at the core of almost all QCD predictions at colliders with hadrons in the initial state,
these logarithms are enhanced in the $\overline{\rm MS}$ scheme
both in the partonic coefficient functions and the splitting functions governing DGLAP evolution.
Their resummation is possible thanks to the work of several groups~\cite{Salam:1998tj,Ciafaloni:1999yw,Ciafaloni:2003kd,Ciafaloni:2003rd,Ciafaloni:2007gf,Ball:1995vc,Ball:1997vf,Altarelli:2001ji,Altarelli:2003hk,Altarelli:2005ni,Altarelli:2008aj,Thorne:1999sg,Thorne:1999rb,Thorne:2001nr,White:2006yh,Rothstein:2016bsq, Catani:1990xk,Catani:1990eg,Catani:1994sq, Bonvini:2016wki,Bonvini:2017ogt,Bonvini:2018iwt,Bonvini:2018xvt}.

\begin{figure}[tb]
\centerline{%
\includegraphics[height=4.9cm,page=1]{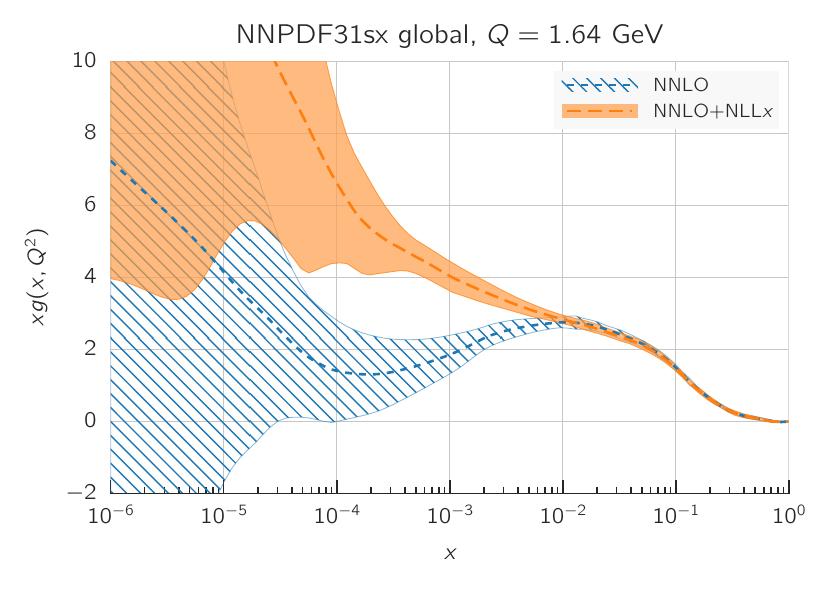}
\includegraphics[height=4.9cm,page=1]{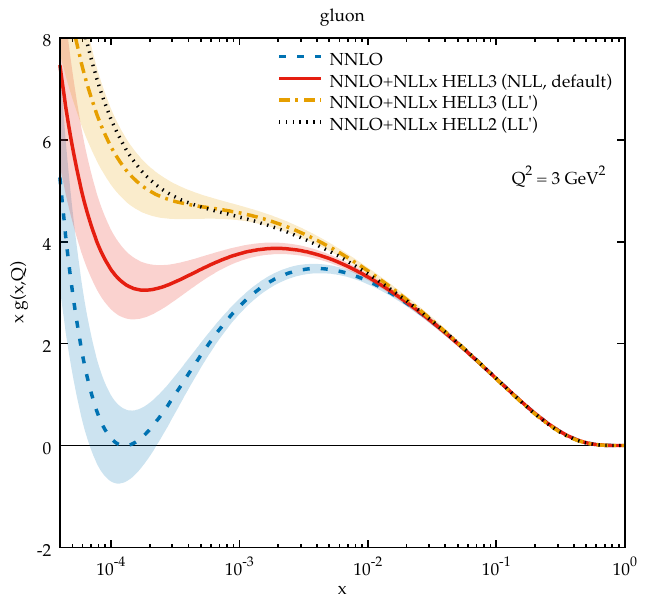}}
\caption{The gluon PDF obtained from a fixed-order fit and from a fit with resummation.
  The left plot show the NNPDF determination~\cite{Ball:2017otu} from a global dataset.
  The right plot shows a determination from HERA data only~\cite{Bonvini:2019wxf}
  and the comparison of different choices of subleading contributions in the resummed prediction.
}
\label{Fig:gluon}
\end{figure}

Recently, small-$x$ resummation has been included in the theory description of PDF evolution
and DIS computations to determine PDFs from data~\cite{Ball:2017otu,Abdolmaleki:2018jln,Bonvini:2019wxf}.
As a result, the fit quality improves substantially, and the PDFs change,
in particular the gluon PDF which is much harder at small $x$.
This is shown in figure~\ref{Fig:gluon}.
The right plot also shows the difference obtained in the resummed gluon (red and orange curves)
when varying the resummation by unknown subleading contributions, as described in Ref.~\cite{Bonvini:2019wxf}.
Indeed, the resummation is currently accurate only to a rather low logarithmic order,
namely leading log (LL) $\as^k\log^k\frac1x$ and next-to-leading log (NLL) $\as^{k+1}\log^k\frac1x$,
for all $k$.
In fact, most ingredients (e.g.\ DIS coefficient functions) are zero at LL, and thus only the first
nontrivial order is known.
For this reason, subleading logarithmic contributions are important and, as one can see from the plot,
varying them may lead to significant differences.

This observation motivates further work towards the extension of resummation to the next logarithmic order.
This is particularly important for the resummation of coefficient functions of processes entering PDF fits,
which are based on the interplay of collinear factorization
with the $k_t$ factorization~\cite{Catani:1990xk,Catani:1990eg,Catani:1994sq},
currently established only at the lowest nontrivial order.

It is worth stressing that current PDF fits with resummation use fixed-order theory at NNLO,
the present state-of-the-art accuracy for PDF determination.
However, there is an increasing activity in the PDF fitting community to
go beyond NNLO, see e.g.\ the recent approximate N$^3$LO fit by the MSHT collaboration~\cite{McGowan:2022nag}.
One important aspect to keep in mind is that the instability at small $x$ induced by the
presence of small-$x$ logarithms at fixed order is more severe at N$^3$LO than at NNLO~\cite{Bonvini:2018xvt}.
Therefore, in order to achieve a reliable description of the small-$x$ region
in PDF fits using N$^3$LO theory, the inclusion of small-$x$ resummation will be fundamental.

\section{Impact of resummation at LHC and FCC}

\begin{figure}[tb]
\centerline{%
\includegraphics[width=6.2cm,page=1]{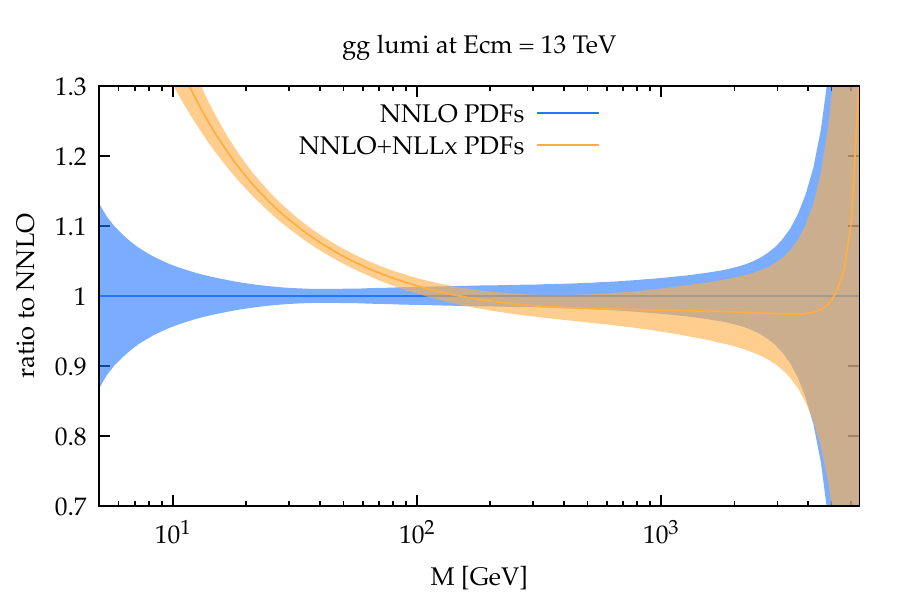}
\includegraphics[width=6.2cm,page=2]{images/PDF_lumi_LHC}}
\centerline{%
\includegraphics[width=6.2cm,page=3]{images/PDF_lumi_LHC}
\includegraphics[width=6.2cm,page=4]{images/PDF_lumi_LHC}
}
\caption{Gluon-gluon (left) and quark-gluon (right) luminosities at LHC $13$~TeV
  computed from the PDF sets of Ref.~\cite{Ball:2017otu}
  obtained with and without the inclusion of small-$x$ resummation.
  The upper plots show the integrated luminosity as a function of the invariant mass,
  the lower plots show the differential luminosity as a function of rapidity for fixed invariant mass $M=30$~GeV.}
\label{Fig:LHC}
\end{figure}

\begin{figure}[tb]
\centerline{%
\includegraphics[width=6.2cm,page=1]{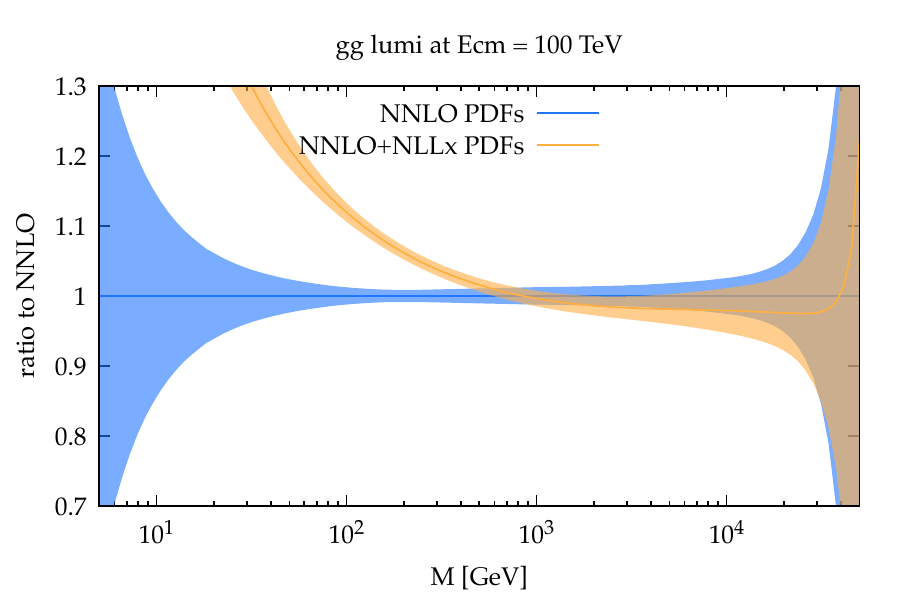}
\includegraphics[width=6.2cm,page=2]{images/PDF_lumi_FCC}}
\centerline{%
\includegraphics[width=6.2cm,page=3]{images/PDF_lumi_FCC}
\includegraphics[width=6.2cm,page=4]{images/PDF_lumi_FCC}
}
\caption{Same as Fig.~\ref{Fig:LHC} but for a future collider of $100$~TeV energy.
  The lower plots are for invariant mass $M=100$~GeV.}
\label{Fig:FCC}
\end{figure}

The significant effect of small-$x$ resummation in the determination of the gluon PDF
has important implication in the phenomenology at LHC and future colliders.
To appreciate such effect we show parton luminosities in figure~\ref{Fig:LHC} for the LHC
and in figure~\ref{Fig:FCC} for a future collider (FCC) of energy $100$~TeV.
At the LHC, we see that, for invariant masses below approximately $100$~GeV,
the $gg$ and $qg$ luminosities are larger at resummed level than at fixed order,
reaching an increase of a few tens of percent at masses of the order of $10$~GeV.
At the FCC the situation is similar, but the effect of resummation starts at invariant masses
of about $500$~GeV, and it is therefore larger at smaller scales.
Therefore, while at the LHC resummation is expected to be important for low-scale physics
(below the electroweak scale, e.g.\ for $b$ physics), at higher energy colliders
it becomes a crucial ingredient for the description of electroweak physics,
and in particular for Higgs production which is central for the FCC physics programme.

From the plots differential in rapidity, we note that the increase induced by resummaiton is larger,
in percentage, towards the rapidity endpoint, in particular in the direction of the gluon.
Indeed, at large rapidity one of the two partons is at small $x$, while the other is at large $x$.
This observation has two implications. The first is that the effect of resummation
is more clearly visible when looking at differential distributions.
The second is that a good description of the large rapidity region may require
the interplay of small-$x$ resummation with large-$x$ resummation,
as both regimes contribute to that kinematic region.

A first study of the simultaneous resummation of small- and large-$x$ logarithms
has been considered in Ref.~\cite{Bonvini:2018ixe} for Higgs production at proton-proton colliders.
However, in that case only the total cross section has been considered,
and the effect of the two resummations is additive.
The two resummations are more intimately connected at differential level,
so future work in the direction of resumming both types of logarithms
in differential observables is called for.

\begin{figure}[tb]
\centerline{%
\includegraphics[width=6.2cm,page=3]{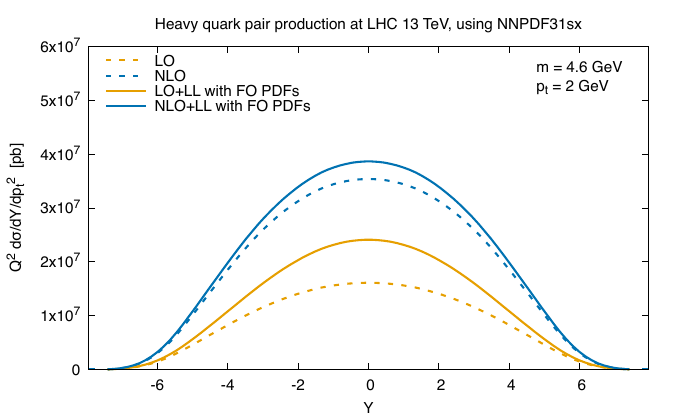}
\includegraphics[width=6.2cm,page=4]{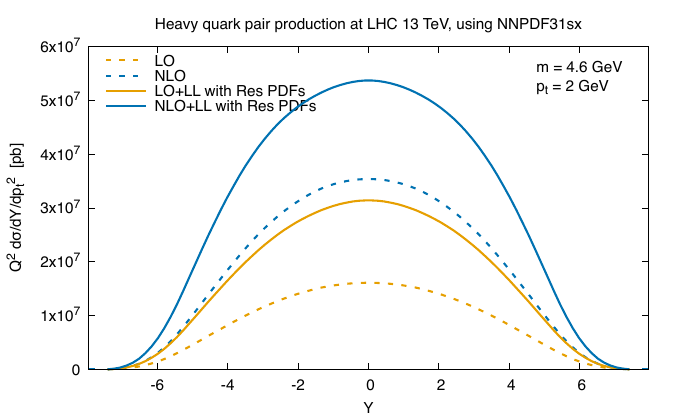}}
\caption{Differential distribution in rapidity and transverse momentum of a bottom quark
  in $b\bar b$ production at LHC, as a function of the rapidity~\cite{Silvetti:2022hyc}.
  In the left plot the same PDFs are used for both fixed-order and resummed results,
  while in the right plot the resummed result uses resummed PDF.}
\label{Fig:HQ}
\end{figure}

As far as small-$x$ resummation is concerned,
recently the formalism introduced in Refs.~\cite{Bonvini:2016wki,Bonvini:2017ogt,Bonvini:2018iwt}
has been extended to differential distributions in Ref.~\cite{Silvetti:2022hyc}.
There, the process of heavy quark pair production has been considered as a representative
application of the formalism.
In figure~\ref{Fig:HQ} we report the result of Ref.~\cite{Silvetti:2022hyc}
for the production of a bottom quark pair at the LHC. In particular the plot shows
the double differential distribution in rapidity and transverse momentum of one of
the produced bottom quarks, plotted as a function of the rapidity and
for fixed transverse momentum $p_t=2$~GeV.
The left plot shows the effect of resummation in the coefficient function only,
as both fixed-order and resummed results are obtained with the same (fixed-order) PDFs.
We see that resummation improves the convergence pattern of the perturbative expansion,
with the resummed curves at two adjacent orders being closer than the analogous results at fixed order.
In the right plot the resummed result is obtained using resummed PDFs,
that further increase the effect of resummation, as expected from figure~\ref{Fig:LHC}.
This result shows that this process has the potential to give additional
important constraints in PDF determination, but to describe it reliably
small-$x$ resummation must be included.

\addcontentsline{toc}{section}{References}
\bibliographystyle{jhep}
\bibliography{references}

\end{document}